\documentclass[12pt]{article}

\usepackage[mathscr]{eucal} 
\usepackage{amscd}
\usepackage{amssymb,latexsym}
\newtheorem{thm}{Theorem}%[section]
\newtheorem{cor}[thm]{Corollary}%[section]
\newcommand{\qed}{\hfill $\Box$ \medskip}
\newcommand{\proof}{\noindent \emph{Proof:\ }}

\renewcommand\l{\lambda}

\renewcommand\S{\Sigma}
\newcommand\s{\sigma}
\renewcommand\d{\partial}

\newcommand\f{\phi}
\renewcommand\L{\triangle}
\newcommand\D{\nabla}
\newcommand\e{\epsilon}

\renewcommand\div{{\rm div}}

\newcommand\la{\langle}
\newcommand\ra{\rangle}

\newcommand\ric{{\rm Ric}}
\renewcommand\l{\lambda}

\newcommand\8{\infty}

\renewcommand\th{\theta}

\newcommand\beq{\begin{eqnarray}}
\newcommand\eeq{\end{eqnarray}}
\newcommand\ben{\begin{enumerate}}
\newcommand\een{\end{enumerate}}

\newcounter{mnotecount}[section]

\title{On the topology and area of higher\\ dimensional black holes}

\author{Mingliang Cai and Gregory J. Galloway\thanks{Supported in part by NSF grant \# DMS-9803566} \\
 Department of Mathematics\\University of Miami\\ Coral Gables FL 33124, USA}
\begin{document}
\maketitle

\begin{abstract}
Over the past decade there has been an increasing interest in the study of black holes,  
and related objects, in higher (and lower) dimensions, motivated to a large extent by developments in string
theory.   The aim of the present paper is to obtain higher dimensional
analogues of some well known results for black holes in $3+1$ dimensions.  More precisely, we obtain extensions
to higher dimensions of Hawking's black hole topology theorem for asymptotically flat ($\Lambda =0$)  black hole
spacetimes, and  Gibbons' and Woolgar's genus dependent, lower entropy bound for topological black holes in
asymptotically locally anti-de Sitter ($\Lambda<0$)  spacetimes.  In higher dimensions the genus is replaced by
the so-called  
$\sigma$-constant, or Yamabe invariant, which is a fundamental topological invariant of smooth
compact manifolds.

\end{abstract}

\section{Introduction}

Among the most fundamental macroscopic features of a black hole are its topology and area. 
The first general theorem on the topology of black holes was due to Hawking \cite{H1} in the early 70's  
who proved that the boundary surface of a black hole (obtained by intersecting the event horizon with a suitable
spacelike hypersurface) in a $3+1$ dimensional asymptotically flat stationary black hole spacetime obeying the
dominant energy condition is  spherical.  Much later, during the mid-nineties, it was realized that
results on topological
censorship could be used to improve various aspects of Hawking's black hole topology theorem, cf., \cite{GSWW1}
and references cited therein.  At the same time, during the 90's, it was recognized that under appropriate
circumstances, one could have as  black hole boundaries surfaces of higher genus.  As is now
well known, such examples can occur in $3+1$ dimensional black hole spacetimes 
which are asymptotically locally anti-de
Sitter, and  which, in particular, have nontrivial topology at infinity, see e.g., \cite{BLP}, and references
cited therein.  Such examples, of course, do not violate Hawking's theorem; because of the presence of a negative
cosmological constant, the dominant energy condition is not satisified.  These examples, however, do satisfy the
mild energy conditions of topological censorship, and, in fact, as  shown in \cite{GSWW1}, are
consistent with topological censorship.  In the general setting of asymptotically locally anti-de Sitter
spacetimes, topological censorship implies that the genus of the black hole boundary (or, if there is more than
one black hole, the sum of the genera of the black hole boundaries) is bounded above by the genus of the surface at
infinity; in particular, if the latter is spherical, so is the black hole boundary.  
Hence, in $3+1$ dimensions, the
topology of black holes is controlled by the topology at infinity. 

Although Hawking's theorem does not hold in the asymptotically locally anti-de Sitter setting, his
basic argument still applies and leads to an interesting conclusion, as  pointed out by Gibbons
\cite{Gi3} in the time-symmetric case.  In this case,  under circumstances closely related to those considered by
Hawking, and assuming nonnegative local energy density $T_{00} \ge 0$,
Gibbons obtains the following lower bound for the area of a black hole boundary $\S$, 
\beq\label{bound}
{\rm Area}(\S)\ge \frac{4\pi(g-1)}{|\Lambda|} \, ,
\eeq
where $g$ is the genus of $\S$, and $\Lambda < 0$ is the cosmological constant.  Hence, in this situation
the black hole entropy has a lower bound depending on a global topological invariant.  Shortly thereafter,
Woolgar \cite{Wo} obtained independently a similar inequality in the general (nontime-symmetric) case.

During the past decade there has been  a significant increase in interest in black holes in higher (and lower) 
dimensions, due largely to the recognition of their relevance to string theory; see for example the review
articles of Horowitz \cite{Ho} and Peet \cite{Pe}.
There is now an extensive literature of solutions in string theory (and
supergravity) with horizons, which represent black holes, and  related objects, such as black strings
and black $p$-branes, in various dimensions. The physical properties of these solutions have been widely studied.
Interest in black holes in various dimensions has intensified in recent years, due, for example, to the 
role they have played in the conjectured  correspondence between string theory (and supergravity) on
asymptotically locally anti-de Sitter backgrounds and the large $N$ limit of certain conformal field
theories defined on the boundary-at-infinity of these backgrounds, cf. \cite{Ma+,Ma,Wi}.  

The aim of the present paper is to establish some
general properties of black holes which hold in arbitrary dimension.  To be more specific, our goal is
to obtain extensions of Hawking's black hole topology theorem (relevant when $\Lambda \ge 0$) and 
Gibbon's entropy bound (relevant when $\Lambda < 0$) to higher dimensions. 
The general setting for these results is presented in the next section.  We restrict
attention primarily to the time-symmetric case.  Some comments regarding the general case are given
in Section~4. 

In Section 2 we obtain restrictions on the topology of a black hole boundary surface $\S^{n-1}$ sitting
in a spacelike hypersurface $V^n$ in a spacetime $M^{n+1}$, $n\ge 3$.  We show under suitable circumstances,
(in particular, under appropriate energy conditions) that $\S^{n-1}$ must admit a metric of
positive scalar curvature (which needn't be 
the induced metric).  In the $3+1$ ($\dim \S = 2$) case, this implies, by 
Gauss-Bonnet that
$\S$ is a
$2$-sphere, and we recover Hawking's theorem.  The point, however, is that there are many known topological
obstructions to metrics of positive scalar curvature in higher dimensions, as well, as we recall in Section 2.
In particular, in the $4+1$ ($\dim \S =3$) case, the topology  of $\S$  is
restricted to the $3$-sphere (modulo the Poincar\'e conjecture) perhaps with identifications,  $S^2\times S^1$,
and finite connected sums of such manifolds.  

The proofs of the results in Section~2 are based on a variation of 
Hawking's original argument.  However, his argument does not directly extend to higher dimensions, since, in higher
dimensions, the Gauss-Bonnet theorem cannot be used, as it was by Hawking, to control the total scalar curvature of
$\S$.   
Instead, we combine Hawking's argument together with arguments similar to those of Schoen  and Yau \cite{SY2} in their study
of higher dimensional manifolds of positive scalar curvature. 
We mention also, that, although topological censorship holds in arbitrary dimension, the techniques
used in \cite{GSWW1} to study the topology of black holes in $3+1$ dimensions  via topological censorship do
not directly extend to higher dimensions, see \cite{GSWW2} for further discussion of this point.  

In Section 3 we obtain a higher dimensional analogue of the entropy bound (\ref{bound}).  
Given the role that scalar curvature plays in Hawking's argument, and in the
higher dimensional arguments of Section 2, one is led to consider the
so-called \emph{$\s$-constant} (or Yamabe invariant) \cite{An1,Sch} as the relevant topological invariant 
in higher dimensions.   
With regard to this choice we were also influenced by suggestive  comments of Gibbons~\cite{Gi3}.
The $\s$-constant is a fundamental topological invariant of smooth compact manifolds, 
which we now briefly describe; see \cite{An1} for further details.

Let $\S^{n-1}$, $n\ge 3$, be a smooth compact (without boundary) $(n-1)$-dimensional
manifold.  If $g$ is a Riemannian metric on $\S^{n-1}$, let $[g]$ denote the conformal
class of $g$.  The Yamabe constant with respect to $[g]$, which we denote by $\lambda[g]$, is the number,
\beq\label{yam}
\l[g]  = \inf_{\tilde g\in [g]} 
\frac{\int_{\S}S_{\tilde g}d\mu_{\tilde g}}
{(\int_{\S}d\mu_{\tilde g})^{\frac{n-3}{n-1}}}\, ,
\eeq  
where $S_{\tilde g}$ and $d\mu_{\tilde g}$ are respectively the scalar curvature and volume measure of $\S^{n-1}$ 
in the metric $\tilde g$.  The  expression involving integrals is just the volume-normalized total
scalar curvature of $(\S,\tilde g)$.
The solution to the famous Yamabe problem, due to Yamabe, Trudinger, Aubin and Schoen, 
guarantees that the infimum 
in (\ref{yam}) is  achieved by a metric of constant scalar curvature.  

The $\s$-constant  of $\S$ is 
 defined by taking the supremum of the Yamabe constants over all conformal
classes,
\beq
\s(\S) = \sup_{[g]} \l[g] \, .
\eeq    
As observed by Aubin, the supremum is finite, and in fact bounded above in terms of the volume
of the standard unit $(n-1)$-sphere $S^{n-1} \subset \Bbb R^n$.  The $\s$-constant divides
the family of compact manifolds into three classes according to: (1) $\s(\S) > 0$, (2) $\s(\S) = 0$,
and (3) $\s(\S) < 0$.  It follows from the solution of the Yamabe problem that $\s(\S) > 0$ if and only if
$\S$ admits a metric of positive scalar curvature.  

In the case $\dim \S =2$, the Gauss-Bonnet theorem implies $\s(\S) = 4\pi\chi(\S)=8\pi(1-g)$.  In a 
certain sense, then, one may view  
the $\s$-constant as a generalization of the Euler characteristic to higher dimensions. 
This has been  especially emphasized in three dimensions by Anderson, who in \cite{An1} describes
some fundamental connections between the $\s$-constant and 
Thurston's geometrization conjecture for
$3$-manifolds. In this context there are some well known results and conjectures concerning the sign of the
$\s$-constant.  As noted above, $\s(\S) \le 0$ if and only if $\S$ does not carry a metric of positive scalar
curvature; some large classes of examples for which this holds are discussed in Section 2.  It is believed
that all compact  hyperbolic $3$-manifolds (which, in fact, accounts for most compact $3$-manifolds) have strictly
negative $\s$-constant, but this has yet to be proven.
There has  been
progress in determining the sign of the $\s$-constant in $4$-dimensions; see for example the paper of Lebrun
\cite{Le}, in which Seiberg-Witten theory is used to determine the sign of the $\s$-constant within the class
of compact complex algebraic surfaces (of, hence, four real dimensions) in terms of the Kodaira dimension. 
Finally, we mention that the $\s$-constant has also arisen in the cosmological 
context, see for example, \cite{An2, FM}; related notions have appeared in some earlier papers, 
as well, cf., \cite{Gi2,O'M}.  

Let $\S^{n-1}$ be a smooth black hole boundary 
contained in a spacelike hypersurface $V^n$ in a spacetime $M^{n+1}$,
as described in Section 2, and suppose $\s(\S)\le 0$ (which, in higher dimensions, is roughly analogous to the
assumption
$g\ge 1$ in the $\dim \S =2$ case).
Then the results in Section 3 imply, under appropriate circumstances
(e.g., $T_{00} \ge 0$ and $\Lambda < 0$), that the area of $\S^{n-1}$ satisfies,
\beq\label{vol2}
{\rm vol}(\S^{n-1}) \ge  \left|\frac{\s(\S)}{2\Lambda}\right|^{\frac{n-1}2} \, ,
\eeq
in analogy with the entropy bound~(\ref{bound}).

We  proceed to a detailed presentation of our results.  Some concluding remarks are made
in Section 4. 
     
\section{Topological restrictions}

We now describe the basic setting for our results.
Let $\tilde V^n$ be a smooth spacelike hypersurface in a  spacetime $M^{n+1}$, and  let 
$\S^{n-1}$ be a smooth compact hypersurface in $\tilde V^n$. 
To simplify certain statements and arguments, we assume $\S^{n-1}$ is connected.  However, all of the
results presented here apply to each component of $\S^{n-1}$, in cases that it is not connected. 
Physically one is to think of $\S^{n-1}$
as the intersection of $\tilde V^n$ with the black hole event horizon in a stationary or static 
black hole spacetime.  Alternatively,
for dynamic black holes, one may think of
$\S^{n-1}$ as an  apparent horizon contained within the black hole region.  In either case, we assume that
 $\S^{n-1}$ is marginally outer trapped,  i.e., that the null expansion $\theta$ along $\S^{n-1}$, 
with respect to the outward null normals, vanishes.  Implicit in this
assumption is that $\S^{n-1}$ separates $\tilde V^n$ into an ``inside" and an ``outside".
Let $V^n$ denote the outside of $\tilde V^n$ together with $\S^{n-1}$; hence, $V^n$, in the induced metric,
is a smooth Riemannian manifold-with-boundary, with compact boundary $\S^{n-1}$.
Although some of our results will hold in more generality, for simplicity we are going to restrict
attention to the time-symmetric case.  In particular, our results apply in a natural way 
to the hypersurfaces of orthogonality in static black hole spacetimes.  Some comments concerning the 
nontime-symmetric case are presented in Section 4.

Thus, we assume that $V^n$ is a hypersurface of time symmetry,
i.e., is totally geodesic.  Then, as is standard in this case,
conditions concerning the null expansion scalar along a hypersurface $W^{n-1}$ in $V^n$ 
can be expressed in terms of the 
mean curvature of $W^{n-1}$ as a submanifold of $V^n$.  In particular, 
the assumption that $\S^{n-1}$ is marginally outer
trapped, reduces to the assumption that $\S^{n-1}$ is a minimal 
surface (i.e., has vanishing mean curvature, $H=0$) in $V^n$.
Similarly, an outer trapped surface in $V^n$ is a compact hypersurface $W^{n-1}$ in 
$V^n$ homologous to
$\S^{n-1}$ which  has negative mean curvature, $H < 0$, with respect to its outward unit normal in $V^n$.  
(By ``outward normal", we mean the normal to $W^{n-1}$ which points out from the region
bounded by $\S^{n-1}$ and $W^{n-1}$.  By our sign  conventions,
$H = \mathrm{div}_W N$, where $N$ is the outward pointing unit normal to $W^{n-1}$ in $V^n$.)  

We say that $V^n$ is
\emph{regular} if, in addition to $\Sigma^{n-1}$ being minimal, there are no outer trapped surfaces
in the interior of $V^{n}$.  
The nonexistence of outer trapped surfaces in the domain of outer communications of
a black hole spacetime, which we have incorporated into our model, is a standard result of black hole theory as
developed in\cite{HE}.  The arguments involved in establishing this do not depend on the dimension in an
essential way.   In all of
our results it is assumed that $V^n$ is regular (or else satisifies a slightly 
 stronger condition which we describe
later). 

Since $V^n$ is totally geodesic, the scalar curvature $S$ of $V^n$ is geometrically constrained by the
Gauss equations to satisfy
\beq
S = 2 R_{00} + R \, ,
\eeq 
where $R_{00}$ is the Ricci curvature of spacetime in the direction orthogonal to $V^n$ and $R$ is the scalar
curvature of spacetime.  In this paper we are primarily concerned with physical theories on spacetime whose field
equations include equations of the form,
\beq
R_{\mu\nu} +\frac12 Rg_{\mu\nu}+ \Lambda g_{\mu\nu} = T_{\mu\nu}, 
\eeq
where $T_{\mu\nu}$ includes various matter field terms, and $\Lambda$ is constant.  This includes general
relativity, of course, but also  various string field theories at sufficiently low energies, cf., \cite{Ho, Pe}
and references cited therein.   In particular, our results apply to various black string and black $p$-brane
solutions, as well as to conventional higher dimensional black holes.
Taking into account the field equations, $S$ is constrained to satisfy,
\beq
S = 2T_{00} + 2\Lambda \, .
\eeq

Our first result establishes restrictions on the topology of the horizon $\S^{n-1}$ under the assumption
that $S = 2T_{00} + 2\Lambda$ is nonnegative.  (We have in mind, in particular, the 
asymptotically flat case, in which $\Lambda =0$ 
and the energy condition, $T_{00}\ge 0$, holds.) Under this
assumption it is shown that
$\S^{n-1}$ carries  (generically) a metric of
positive scalar curvature.  One may then appeal to the vast literature of results 
establishing restrictions on the
topology of compact manifolds that admit metrics of positive scalar curvature, 
cf., \cite{GL2} for an overview.   Particularly strong results hold in the $4+1$ ($\dim\S=3)$ case.  
We will recall some of these results after the proof of the following theorem. 

\begin{thm}\label{top}
Let $V^{n}$, $n\ge 3$, be a regular time symmetric spacelike hypersurface, with compact boundary horizon 
$\S^{n-1}$, in a spacetime $M^{n+1}$, as described above.  Suppose the scalar curvature of $V^{n}$ satisfies, 
$S = 2T_{00} + 2\Lambda\ge 0$ along $\S^{n-1}$.  Then either
\ben
\item[(a)] $\S^{n-1}$ admits a metric of positive scalar curvature, or 
\item[(b)] $S = 2T_{00} + 2\Lambda \equiv 0$ on $\S^{n-1}$, and $\S^{n-1}$ is totally geodesic and Ricci flat. 
\een
\end{thm} 

In the $3+1$ ($\dim\S =2$) case,  the theorem implies, using Gauss-Bonnet, that $\S^2$ is either a flat torus or a
topological sphere, in agreement with standard results \cite{H1,H2,HE}.  (We recall that the torus
arises as a borderline case in Hawking's original arguments, as well.  The torus case is not easily eliminated
without further assumptions, cf., the discussion in \cite{Ga}.) 

\proof  The proof is similar in spirit to the proof of Hawking's black hole topology theorem in 
$3+1$ dimensions \cite{H1,HE}, and
its extension to apparent horizons \cite{H2}, attributed to Gibbons.  However, these proofs
use the Gauss-Bonnet theorem to control the total scalar curvature of the horizon.  This does not work in higher
dimensions.  Instead, we use a variation of the approach taken by Schoen and Yau  \cite{SY2} in their study  of manifolds
of positive scalar curvature in higher dimensions.       

Let $t\to \S_t$ be a variation of $\S_0=\S$ in $V^n$, with variation vector field $X 
=\phi N$, where
$N$ is the unit normal along $\S$ pointing into $V^n$ and 
$\phi$ is a smooth positive function on $\S$.  For $t\ge 0$ sufficiently small, $\{\S_t\}$ foliates a neighborhood
of  $\S$.  Extend $N$ to be the unit normal field to these $\S_t$'s.   For each~$t$, let $H = H_t$ be the mean
curvature of $\S_t$; by our conventions, $H= \div(N)$.
Then by Raychaudhuri's equation
for a (possibly) nongeodesic unit vector field, which is valid  for Riemannian, as well
as Lorentzian, manifolds of arbitrary dimension, we have,
\beq\label{eq:d}
N(H) = - \mathrm{Ric}(N,N)- |B|^2 +\div(\D_NN) \, ,
\eeq
where $\nabla$ and  $\ric$ are respectively the Levi-Civita connection and  Ricci tensor of $V^n$, and
$B = B_t$ is the second fundamental form of $\S_t$: For $X,Y\in T_p\S_t$, $B(X,Y) = \la\D_XN,Y\ra$.

The Gauss equations may be used to obtain the following ``rearrangement" \cite{Gi1,SY1,SY2} 
of the first two terms on the right hand side of (\ref{eq:d}),
\beq\label{eq:e} 
\ric(N,N) + |B|^2 = -\frac12\hat S+\frac12  S + \frac12|B|^2  \quad\mbox{along }\S\, , 
\eeq
where $\hat S$ is the scalar curvature of $\S$ in the induced metric, and we have used the fact that
$\S$ is minimal ($H=0$).  

The divergence term in (\ref{eq:d}) may be written as,
\beq\label{eq:f}
\div(\D_NN) = \div_{\S_t}(\D_NN) - |\D_NN|^2.
\eeq
A further computation shows,
\beq\label{eq:g}
\D_NN = -\frac1{\phi}\D\f \quad\mbox{along } \S \, ,
\eeq
where $\D \f={\rm grad}_{\S} \phi$. 
Substitution of (\ref{eq:g}) into (\ref{eq:f}) gives,
\beq\label{eq:h}
\div(\D_NN) = -\frac1{\phi}\triangle \f \quad \mbox{along } \S \, ,
\eeq
where $\triangle =\triangle_{\S}$ is the Laplacian on $\S$.

By substituting (\ref{eq:e}) and (\ref{eq:h}) into (\ref{eq:d}),  and noting that along $\S$, $\frac{\d H}{\d t} = X(H)
=\f N(H)$, we obtain,
\beq\label{eq:i}
\frac{\d H}{\d t} & = & - \L\f + \frac12(\hat S - S - |B|^2)\f \,\nonumber\\ 
& = & L(\f) \quad\mbox{along }\S \, ,
\eeq
where $L = -\L + \frac12(\hat S - S - |B|^2)$ is the so called \emph{stability operator},
rearranged according to (\ref{eq:e}). 
 
Let $\l_1$ denote the first eigenvalue of $L$, and let $\f$ be an associated eigenfunction,
\beq\label{eq:j}
L(\f) = \l_1\f \, .
\eeq
It is well known that for operators of the form of $L$,  $\f$ can be chosen to be strictly
positive, $\f>0$.  

We observe that $\l_1$ cannot be negative; otherwise (\ref{eq:i}) and (\ref{eq:j}) would imply that
$\frac{\d H}{\d t} <0$ along $\S$.  Since $H=0$ along $\S$, this would mean that for $t$ sufficiently small,
$\S_t$ would be outer trapped, contrary to our assumptions.

Hence, $\l_1 \ge 0$.  
This implies~\cite{FCS} that 
$\S$ is a \emph{stable} minimal hypersurface,
by which we mean that $A''(0) \ge 0$ with respect to every variation $t\to \S_t$ of $\S_0 =\S$,   
where $A(t) = $ the area of $\S_t$.  The conclusion of Theorem \ref{top} then follows from 
arguments in \cite{SY2}.  Here we give a more direct argument, which is also relevant to the
nontime-symmetric case.
Let $g$ denote the induced metric on $\S$, and consider the conformally related metric
$\tilde g = \f^{\frac2{n-2}} g$.  
The scalar curvature $\tilde S$ of $\S$ in the metric
$\tilde g$ is given by,

\beq\label{eq:k}
\tilde S & = & \phi^{-\frac{n}{n-2}}(-2\L\f + \hat S\f +\frac{n-1}{n-2}\frac{|\D\f|^2}{\f}) \nonumber \\  
& = & \phi^{-\frac{2}{n-2}}(2\l_1 + S + |B|^2 +\frac{n-1}{n-2}\frac{|\D\f|^2}{\f^2})
\eeq 
where, for the second equation, we have used  (\ref{eq:j}).

Equation (\ref{eq:k}) and the assumption $S\ge 0$ imply that $\tilde S\ge 0$.  If $\tilde S > 0$ at some point, then by
well known results of Kazdan and Warner \cite{KW}, one can conformally change $\tilde g$ (and, hence $g$) to a
metric of strictly positive scalar curvature. Case~(a) of Theorem \ref{top} then holds.  If $\tilde S$ vanishes
identically, then, by equation (\ref{eq:k}),  $\l_1 = 0$, $S \equiv 0$ along $\S$, $B= 0$, and $\f$ is
constant.    Equation (\ref{eq:j}) then implies that $\hat S \equiv 0$. By a result of Bourguinon  (see 
\cite{KW}), it follows that $\S$ carries a metric of 
positive scalar curvature, unless it is Ricci flat.  Theorem~\ref{top} now  follows.
\qed

\noindent
{\it Remark:} As shown in the proof,  the absence of  outer trapped surfaces 
in $V$ implies that $\S$ is stable.
We make use of this observation, which  is valid in the time-symmetric case only, 
in Section 3.  In Section 4 we consider the extension of Theorem \ref{top} to the nontime-symmetric 
case.

\vspace{1ex}

Theorem \ref{top} asserts that, apart from exceptional cases, $\S^{n-1}$ carries a metric of positive scalar curvature.
(In fact, later we will show, under slightly stronger conditions, that case (b) cannot occur.)
Although the scalar curvature is a rather weak geometric invariant, there are  
many known topological obstructions to the existence of metrics of positive scalar curvature.
In the celebrated paper of Lichnerowicz ~\cite{Li}, spinorial methods and the Atiyah-Singer index
theorem were used to prove that compact $4k$-dimensional spin manifolds of positive scalar curvature have
vanishing $\hat A$-genus.  This was followed later by work of Hitchin \cite{Hi}, who also used the spinorial
method to obtain further vanishing theorems.  However these results left open the question as to whether, for
example, the $k$-torus admits a metric of positive scalar curvature. Then in \cite{SY1}, Schoen and Yau proved
that the fundamental group of a compact orientable $3$-manifold that  admits a metric of positive scalar
curvature cannot contain a subgroup isomorphic to a nontrivial surface group.  This implies, in particular, that
the $3$-torus does not admit a metric of positive scalar curvature.   In \cite{SY2}, Schoen and Yau generalized
their techniques to higher dimensions, thereby establishing the existence of a large class of compact manifolds,
including tori, of dimension up to
$7$, that do not admit metrics of positive scalar curvature.  These results were put in a somewhat broader
context, in the case of spin manifolds, by Gromov and Lawson \cite{GL1,GL2} via the development of 
the notion of \emph{enlargability}.  We mention just two results obtained in \cite{GL2}.  It is 
proved there, for example, that any compact manifold (of arbitrary dimension) that admits a metric of positive
scalar curvature, cannot carry a metric of nonpositive sectional curvature $K\le 0$.  This result rules
out many obvious topologies, such as products of tori and higher genus surfaces, etc.  

Now restrict attention to the case $\dim \S = 3$, and assume, by passing to a double cover if necessary,
that $\S$ is orientable.   
According to the prime decomposition theorem  of Milnor, any such manifold $\S$ can be expressed
as a finite connected sum of three types of manifolds:  (i) manifolds  covered by homotopy
$3$-spheres (or $3$-spheres, if the Poincar\'e conjecture holds), (ii) manifolds diffeomorphic to
$S^2\times S^1$, and (iii) $K(\pi,1)$ manifolds.  Recall, a $K(\pi,1)$ manifold is a manifold  whose
universal cover is contractible (such as the torus).  In \cite{GL2} it is shown that if $\S$ admits
a metric of positive scalar curvature then it cannot have any $K(\pi,1)$'s in its prime decomposition.

We are led to the following corollary.

\begin{cor}\label{cor}
Let $V^{4}$ be a regular time symmetric spacelike hypersurface, with compact boundary horizon 
$\S^{3}$, in a spacetime $M^{4+1}$.  Suppose the scalar curvature of $V^{4}$
satisfies,
$S = 2T_{00} + 2\Lambda\ge 0$ along $\S^{3}$.  Then, unless it is flat, $\S^3$ (or a double cover, if
$\S^3$ is nonorientable) is diffeomorphic to a finite connected sum of  homotopy
$3$-spheres, perhaps with identifications, and  $k\ge 0$ copies of $S^2\times S^1$.
\end{cor} 

Thus, the basic black hole topologies for the  $4+1$ dimensional black hole spacetimes modeled in 
Corollary \ref{cor} are
$S^3$ (perhaps with identifications) and $S^2\times S^1$.  If $\S$ is flat then it must be a
$3$-torus, or be covered by a $3$-torus.
The corollary follows immediately from Theorem \ref{top}, the result of Gromov and Lawson  described above,
and the fact that in three dimensions, Ricci flat implies flat.

As mentioned earlier, case (b) of Theorem \ref{top} can be eliminated under slightly stronger conditions.
Let $V^n$ be an orientable time-symmetric spacelike hypersurface with compact minimal ($H=0$) boundary horizon
$\S^{n-1}$.  A marginally outer trapped surface in $V^n$ is a compact minimal hypersurface in 
$V^n$ homologous to $\S^{n-1}$.  We say that $V^n$ is \emph{strongly regular} provided (1) there are
no marginally outer trapped surfaces in the interior of $V^n$, and (2) there exists a compact
hypersurface $\S_0^{n-1}$ in $V^n$ homologous to
$\S^{n-1}$ which is mean convex, i.e., which satisifies $H >0$ with respect to the 
outward normal.  Condition (2) is a
mild  asymptotic condition generally satisfied by  asymptotically flat and 
asymptotically locally anti-de Sitter  
spacetimes.  The nonexistence of marginally outer trapped surfaces in the domain of outer 
communications is a standard
result of black hole theory  as developed in \cite{HE}. 

\newpage
\begin{thm}\label{top2}
Let $V^{n}$, $3\le n\le 7$, be a strongly regular time symmetric spacelike hypersurface, 
with compact boundary horizon 
$\S^{n-1}$, in a spacetime $M^{n+1}$, as described above.  Suppose the scalar curvature of $V^{n}$ satisfies, 
$S = 2T_{00} + 2\Lambda\ge 0$.  Then $\S^{n-1}$ admits a metric of positive scalar curvature.
\end{thm} 

\proof  
Let $V_0$ be the region in $V$ bounded by $\S$ and $\S_0$.  $V_0$ is a compact Riemannian manifold-with-boundary,
with (weakly) mean convex boundary $\d V_0 = \S\cup \S_0$.  $\S$ determines a nontrivial homology class in
$H_{n-1}(V_0,\Bbb Z)$.  We claim that $\S$ is of least area in its homology class.  To see this, we minimize area
in the homology class of $\S$. By standard results in geometric measure theory, there exists a smooth minimizer
contained in $V_0\setminus \S_0$; for the regularity of the minimizer we are using the assumption $n\le 7$.  Let 
$\S_1$ denote a component of this minimizer; $\S_1$ is necessarily minimal, i.e., has mean curvature $H=0$.  To
avoid having a marginally  outer trapped surface contained in the interior of $V$, $\S_1$ must meet $\S$.  
But then, since $\S$ is also minimal, it follows from the maximum principle  that $\S_1 = \S$, which
implies our claim.  

In particular, $\S$ is locally of least area, i.e., the area of $\S$ is less than 
or equal to the area of any nearby
surface isotopic to it.  If the scalar curvature $S$ of $V$ were strictly positive, then  arguments
in \cite{SY2} would imply that $\S$, being locally of least area, would carry a metric of positive 
scalar curvature. 
If, as is our situation, the scalar curvature $S$ is merely assumed to be nonnegative, it is shown in
\cite{Cai} (which generalizes to higher dimension results in \cite{CG}) that $\S$ can fail to carry a
metric of positive scalar curvature only under special circumstances: $\S$ must be totally geodesic, and
a neighborhood $U$ of $\S$ must be isometric to $[0,\e) \times \S$.  Thus, to avoid the occurence
of marginally outer trapped surfaces in the interior of $V$, we conclude that $\S$ carries a metric of
positive scalar curvature. 
\qed 

\noindent
\emph{Remark:}  The dimension restriction $n\le 7$ is most likely an artifact of the proof. It only comes up
in showing that $\S$ is locally of least area; if this is
known to be the case, the dimension restriction is unnecessary.  It is possible that the absence of  
outer trapped and
marginally outer trapped surfaces alone is sufficient to imply that $\S$ is locally of least area, 
but this does not
follow immediately.

\medskip
In the $4+1$ ($\dim \S =3$) case, we obtain the following corollary to Theorem~\ref{top2},
similar to Corollary \ref{cor}.

\begin{cor}\label{cor2}
Let $V^{4}$ be a strongly regular time symmetric spacelike hypersurface, with compact boundary horizon 
$\S^{3}$, in a spacetime $M^{4+1}$.  Suppose the scalar curvature of $V^{4}$
satisfies,
$S = 2T_{00} + 2\Lambda\ge 0$.  Then,  $\S^3$ (or a double cover, if
$\S^3$ is nonorientable) is diffeomorphic to a finite connected sum of  homotopy
$3$-spheres, perhaps with identifications, and  $k\ge 0$ copies of $S^2\times S^1$.
\end{cor}

\section{Lower area bounds and the $\s$-constant}

With the setting as in the previous section, we obtain lower bounds for the area of the horizon $\S^{n-1}$ 
in terms of its $\s$-constant, which is assumed to be nonpositive, $\s(\S)\le 0$, under the
assumption that $S= 2T_{00} + 2\Lambda$ satisfies, $S\ge -\kappa$, where
$\kappa$ is a positive constant.  We have in mind, in particular, the asymptotically locally
anti-de Sitter case, in which  $\Lambda < 0$
and the energy condition $T_{00} \ge 0$ holds.  

\begin{thm}\label{sigma}
Let $V^{n}$, $n\ge 4$, be a regular time symmetric spacelike hypersurface, 
with compact boundary horizon 
$\S^{n-1}$, in a spacetime $M^{n+1}$, as in Theorem~\ref{top}, such that $\s(\S) \le 0$.  Suppose that 
the scalar curvature of $V^{n}$ satisfies, 
$S = 2T_{00} + 2\Lambda\ge -\kappa$, where $\kappa$ is a positive constant.  
Then the area of $\S^{n-1}$, ${\rm vol}(\S^{n-1})$, satisfies,
\beq\label{vol}
{\rm vol}(\S^{n-1}) \ge \left(\frac{|\l[g]|}{\kappa}\right)^{\frac{n-1}2} \ge
\left(\frac{|\s(\S)|}{\kappa}\right)^{\frac{n-1}2} \, ,
\eeq   
where  $g$ is the induced metric on $\S^{n-1}$. 
\end{thm}

In particular, if $T_{00} \ge 0$ and $\Lambda <0$, 
the area of $\S^{n-1}$ satisfies the inequality (\ref{vol2}).

\vspace{1ex}
\proof  By the remark following the proof of Theorem \ref{top}, $\S$ is a stable
minimal hypersurface, i.e., $A''(0) \ge 0$ with respect to every variation $t\to \S_t$ of $\S_0 =\S$, 
where $A(t) = $ the area of $\S_t$.  We recall the formula for the second
variation of area with respect to variations $t\to \S_t$ having variation vector field $X=\phi N$, where $N$ is 
the unit normal to $\S$ pointing into $V$ and $\phi\in C^{\8}(\S)$, 
\beq\label{var}
A''(0) = \int_{\S} (|\D\phi|^2 - (\ric(N,N)+|B|^2)\phi^2) \, d\mu \, .
\eeq
 
Theorem \ref{sigma} is now a consequence of the following purely Riemannian result, which 
may be of some independent interest.  

\begin{thm}\label{sigma2}
Let $V^n$, $n\ge 4$, be a Riemannian manifold with scalar curvature $S$ 
satisfying, $S\ge -\kappa$, where $\kappa$
is a positive constant.  Let $\S^{n-1}$ be a two-sided compact stable minimal hypersurface in $V^n$ with
$\s$-constant $\s(\S) \le 0$.  Then the area of $\S^{n-1}$, ${\rm vol}(\S^{n-1})$, satisfies
the inequalities in (\ref{vol}).
\end{thm}

\proof  The second inequality in (\ref{vol}) is immediate from the definition of the $\s$-constant.  To prove
the first inequality we use the following reformulation of the Yamabe constant \cite{Besse}, 

\beq\label{yam2}
\l[g] = \inf_{\f\in C^{\8}(\S), \f>0} \frac{\int_{\S} (\frac{4(n-2)}{n-3}|\D\f|^2 +\hat S\f^2)\, d\mu}
{(\int_{\S}\f^{\frac{2(n-1)}{n-3}}\, d\mu)^{\frac{n-3}{n-1}}} \, .
\eeq

In view of equations (\ref{eq:e}) and  (\ref{var}) the stability of $\S^{n-1}$ implies that the following 
inequality holds for all
$\f\in C^{\8}(\S)$ (we may assume $\phi >0$),
\beq\label{stab}
\int_{\S} (|\D \f|^2 + \frac12(\hat S-S-|B|^2)\f^2)\, d\mu \ge 0 \, ,
\eeq
and hence,
\beq
\int_{\S} (2|\D \f|^2 +\hat S \f^2) \, d\mu \ge \int_{\S} S\f^2\,d\mu \, .
\eeq
Then, noting that  $2 < \frac{4(n-2)}{n-3}$, we obtain,
\beq\label{kappa}
\int_{\S} (\frac{4(n-2)}{n-3}|\D \f|^2 +\hat S \f^2) \, d\mu & \ge & \int_{\S}
S\f^2\,d\mu
\nonumber\\
& \ge & - \kappa \int_{\S} \f^2\,d\mu \, .
\eeq
By H\"older's inequality,
\beq
\int_{\S} \f^2\,d\mu \le \left(\int_{\S} \f^{\frac{2(n-1)}{n-3}}\, d\mu\right)^{\frac{n-3}{n-1}}
\left(\int_{\S} 1 \, d\mu\right)^{\frac2{n-1}}  \, ,
\eeq
which, when combined with  (\ref{kappa}), gives,
\beq
\frac{\int_{\S} (\frac{4(n-2)}{n-3}|\D\f|^2 + \hat S\f^2)\, d\mu}
{(\int_{\S}\f^{\frac{2(n-1)}{n-3}}\, d\mu)^{\frac{n-3}{n-1}}} \ge - \kappa
\,(\mbox{vol($\S$)})^{\frac2{n-1}} \, .
\eeq
Making use of this inequality in (\ref{yam2}) yields,
$$
\l[g] \ge - \kappa\, (\mbox{vol($\S$)})^{\frac2{n-1}} \, ,
$$
from which the first inequality in Theorem \ref{sigma2} follows.
\qed

Theorem \ref{sigma2} is a higher dimensional analogue of Theorem 3 in \cite{SZ}, as
well as the result of Gibbons discussed in the introduction (in which the
stability of the horizon was assumed).

\section{Final remarks}

Consider the setting of Theorem \ref{top}, in the general, nontime-symmetric case.  Let $t\to \S_t$ be the
variation of $\S$ in the proof of the theorem.  Let $U$ be the timelike future pointing unit normal along
$V$.  Then $K = U+ N$ is a null vector field along $V$ such that $\th=\th_t = \div_{\S_t}K$ is the null
expansion scalar along $\S_t$.   By a similar computation to that given in the proof of Theorem
\ref{top}, one obtains
the following generalization of Equation  \ref{eq:i},
\beq\label{evolve} 
\frac{\d\th}{\d t} = -\L\phi + \frac12(\hat S -2T_{ab}U^aK^b - 2\Lambda-\Theta^2)\phi + 
 \phi\div_{\S}(\D_NU)^P \nonumber \\ +\phi|\D_NN|^2-\phi|\D_NN+ (\D_NU)^P|^2  \, ,
\eeq 
along $\S$, where we have used the assumption that  $\S$ is marginally outer trapped, $\theta
= 0$ along $\S$.  Here, $\Theta$ is the null second fundamental form of $\S$ 
with respect to $K$, and $X^P$ denotes
projection of the vector $X$ onto $\S$.   
In the $3+1$ dimensional case, Hawking's arguments can cope with the ``bad terms" in the above, e.g., the
divergence term gets integrated away; our method in higher dimensions cannot deal so easily  with these terms. 
However, if the term $(\D_NU)^P$ vanishes along $\S$, Equation~(\ref{evolve})
reduces to a form that permits the arguments in the proof of Theorem~\ref{top} to go through.  Thus, whenever this
term vanishes, e.g., if the second fundamental form of $V$ vanishes along
$\S$, or, more generally, if 
$\D_NU \propto N$ along $\S$, then Theorem~\ref{top} generalizes to the nontime-symmetric case.

Finally, we remark that the arguments presented in this paper, as in the case of the standard $3+1$ results, as
well,  require the black hole boundary
$\S^{n-1}$ to be sufficiently smooth, $C^2$, say.  Horizons, however, need not have this degree of regularity, in
general. For issues regarding the regularity of horizons, see for example \cite{Ch+}, and references cited therein. 

We wish to thank Bill Minicozzi for some helpful comments.

%\newpage

\small{

\providecommand{\bysame}{\leavevmode\hbox to3em{\hrulefill}\thinspace}

}

\end{document}